\documentclass[fleqn, usenatbib, x11names]{mnras}

\usepackage{newtxtext,newtxmath}
\usepackage{caption}
\usepackage{subcaption}
\usepackage{multirow}

\usepackage[T1]{fontenc}

\DeclareRobustCommand{\VAN}[3]{#2}
\let\VANthebibliography\thebibliography
\def\thebibliography{\DeclareRobustCommand{\VAN}[3]{##3}\VANthebibliography}

\usepackage{graphicx}	
\usepackage{amsmath}	
\usepackage{orcidlink}
\usepackage{tabularray}
\UseTblrLibrary{booktabs}
\graphicspath{{./}{figures/}}
\usepackage{bm}

\newcommand*{\mailto}[1]{\href{mailto:#1}{#1}}
\newcommand{\Muv}{M_{\mathrm{UV}}}
\newcommand{\muv}{m_{\mathrm{UV}}}

\newcommand{\Mdot}{\dot{M}}

\newcommand{\Vout}{V_\mathrm{outflow}}
\newcommand{\alphaout}{\alpha_{\mathrm{outflow}}}
\newcommand{\alphastar}{\alpha_{\mathrm{\star}}}

\title[JWST high-$z$ galaxy population synthesis]{Population synthesis and astrophysical inference for high-$z$ JWST galaxies}

\author[T. Driskell et al.]{
Trey Driskell,$^{\orcidlink{0000-0001-9472-7179}1}$\thanks{E-mail: \mailto{gdriskel@usc.edu}}
Ethan O.~Nadler$^{\orcidlink{0000-0002-1182-3825}2,1,3}$,
Andrew Benson,$^{\orcidlink{0000-0001-5501-6008}2}$
and Vera Gluscevic$^{\orcidlink{0000-0002-3589-8637}1}$
\\
$^{1}$Department of Physics $\&$ Astronomy, University of Southern California, Los Angeles, CA, 90007, USA\\
$^{2}$Carnegie Observatories, 813 Santa Barbara Street, Pasadena, CA 91101, USA\\
$^{3}$Department of Astronomy \& Astrophysics, University of California, San Diego, La Jolla, CA 92093, USA\\
}

\pubyear{2024}

\begin{document}
\label{firstpage}
\pagerange{\pageref{firstpage}--\pageref{lastpage}}
\maketitle

\begin{abstract}
Observations of the high-$z$ Universe from JWST have revealed a new population of bright, early galaxies. A robust statistical interpretation of this data requires fast forward models that account for uncertainties in galaxy evolution and incorporate observational systematic effects. We present a probabilistic framework for population synthesis of high-$z$ galaxies and inference of their properties. Our framework is based on the semi-analytic galaxy-formation model \textsc{Galacticus}. To infer the astrophysical parameters governing high-$z$ galaxy evolution, we analyze JWST data from the CEERS and NGDEEP surveys and calculate the likelihood of observing individual objects in apparent magnitude--redshift space, for $z\geq8.5$ galaxy candidates. We include observational selection effects due to limited survey volume and depth, as well as photometric redshift uncertainties. We recover the posterior probability distribution for parameters describing star formation and outflow rates. We place an upper limit on the star formation timescale of $500~\mathrm{Myr}$ at a disk velocity of $50~\mathrm{km\ s}^{-1}$, and we infer a characteristic velocity at which the outflow mass-loading factor is $\sim 1$ of $150^{+280}_{-60}~\mathrm{km\ s}^{-1}$, both at $95\%$ confidence. Marginalizing over our astrophysical model, we find that galaxies in CEERS and NGDEEP data occupy halos with virial masses $10^{10\pm 0.5}~M_{\mathrm{\odot}}$ at $8.5\leq z\leq 12$, at $95\%$ confidence. The star formation timescale preferred by our fit is relatively short compared to typical values at lower redshifts, consistent with previous findings. The modeling and analysis framework presented here can enable systematic tests of high-$z$ galaxies' dust content, initial mass functions, and star formation burstiness in the future.
\end{abstract}




\section{Introduction}\label{sec:intro}

JWST observations have begun to probe a new population of early galaxies. While a number of initial, extremely bright high-$z$ galaxy candidates (e.g., \citealt{Donnan221210126,Labbe220712446,Naidu220709434}) were later found to be low-$z$ interlopers (due to, e.g., contamination from AGN, dust, and nebular-line emission; \citealt{Naidu220802794,Fujimoto221103896,Zavala220801816}), spectroscopic measurements have since confirmed a large population of galaxies at redshifts up to $z\approx 14$ \citep{Curtis-Lake221204658,Fujimoto230109482,Robertson221204480,Carniani240518485,DEugenio240406531,Harikane230406658}. The current JWST sample contains a variety of $z\gtrsim 8$ galaxies identified in surveys including The Cosmic Evolution Early Release Science Survey (CEERS), COSMOS-Web, JADES, The Next Generation Deep Extragalactic Exploratory Public (NGDEEP) Survey, and more (e.g., \citealt{Bouwens221206683,CaseyCOSMOSWeb,bagley2023generation,finkelstein2023complete,Harikane220801612,leung2023ngdeep,Robertson231210033}). Many of these systems are several magnitudes fainter than the brightest high-$z$ galaxies that were the focus of early work, and the completeness of these samples is now well-characterized. The ultraviolet luminosity function (UVLF) inferred from these measurements provides an exciting opportunity for comparisons with different models of cosmology and galaxy formation, beyond what was gleaned from studies focusing on the brightest and highest-redshift galaxy candidates alone.

Observed high-$z$ JWST galaxy populations have been compared to theoretical predictions using three main techniques: empirical models that capture statistical properties of halo and galaxy populations (e.g., \citealt{Boylan-Kolchin220801611,Boylan-Kolchin240710900,Ferrara220800720,Mason220714808,Mirocha220812826,Lovell220810479,Prada230411911,Shen230505679,Iocco240313068,Jespersen240300050}), semi-analytic models of galaxy populations built on halo merger trees (e.g., \citealt{Chen230413890,Kravtsov240504578,Lu240602672,Nusser240218942,Yung230404348}), and hydrodynamic simulations that model galaxy formation and evolution in a cosmological context (e.g., \citealt{Wilkins220209431,Kannan221010066,Katz230903269,Keller221212804,Sun230502713,Ceverino240402537,Shen240615548,Feldmann240702674}). Empirical models are generally faster and more flexible, while hydrodynamic simulations capture more of the physical processes relevant for galaxy formation from first principles; see \cite{Wechsler180403097} for an overview.

Despite this variety of techniques, the implications of high-$z$ JWST galaxy population measurements for galaxy formation physics remain unclear. Understanding the physics that shapes high-$z$ galaxy populations is important because galaxy formation at high redshifts may qualitatively differ from that at low redshifts (e.g., \citealt{Dayal180909136,Dekel230304827}); however, the relevant processes are captured differently in each forward-modeling approach described above. For example, a key quantity in many empirical models is the star formation efficiency, which parameterizes the fraction of baryons converted into stars in a dark matter halo of a given mass. In semi-analytic models, halo growth captured by merger trees determines the growth of a gas reservoir for each halo; in turn, this relates to a star formation \emph{rate}, with free parameters controlling the efficiency of the conversion from gas to stars on the one hand, and gas outflows due to feedback on the other. Similar physics is captured in hydrodynamic simulations by sub-grid parameters relating to the critical density threshold for star formation, and the feedback prescription (e.g., the effective physics of energy transfer from supernovae to the surrounding gas); see \cite{Somerville14122712} and \cite{Crain230917075} for reviews.

Recent studies produced different physical interpretations of JWST data, illustrating the importance of theoretical and observational uncertainties for high-$z$ galaxy population modeling. Initial studies using a handful of high-$z$ galaxy stellar mass estimates from the brightest high-$z$ galaxy candidates in early JWST observations reported tension with $\Lambda$CDM predictions (e.g., \citealt{Boylan-Kolchin220801611}). However, follow-up work that focused on UVLFs (rather than stellar mass functions), with better-characterized statistical uncertainties, found no tension (e.g., \citealt{Prada230411911,McCaffrey230413755}). Subsequent studies have focused on UVLF measurements from more complete JWST surveys. For example, \cite{Lu240602672} found that semi-analytic predictions from the \textsc{GALFORM} model in a standard $\Lambda$CDM cosmology agree with a compilation of JWST UVLFs at $z\lesssim 10$, but underpredict UVLFs at $z\gtrsim 12$, unless previous dust growth prescriptions are updated. Meanwhile, \cite{Yung230404348} compared Santa Cruz semi-analytic model predictions to JWST data, finding that UVLFs at $z\gtrsim 10$ are underpredicted without modeling dust, and that this discrepancy can be reconciled by a top-heavy initial mass function (IMF), or by adding a stochastic component to the predicted UVLFs (see \citealt{Shen230505679,Sun230502713}). Beyond galaxy formation physics, \cite{Shen240615548} used an empirical model to find that nonstandard early dark energy cosmologies are potentially in better agreement with JWST UVLFs than $\Lambda$CDM predictions, while, e.g., \cite{Feldmann240702674} report no such tension, when relying on hydrodynamic simulations for forward-modeling of the galaxy population. These examples are not exhaustive, and specific results can likely be reconciled through comparison studies. Nonetheless, this situation highlights the importance of modeling JWST data probabilistically, accounting for both observational and theoretical systematics, which is the focus of this work. 

Here, we introduce a probabilistic inference framework relying on high-$z$ galaxy population synthesis based on semi-analytic modeling, critically advancing previous work in the following ways. First, we capture key processes in high-$z$ galaxy evolution through a small number of easy-to-interpret parameters: the star formation timescale and its halo-mass dependence, and the strength and mass-dependence of feedback-driven outflows; following \cite{Yung230404348}, we do not explicitly model dust. We vary these parameters using the semi-analytic merger-tree-based model \textsc{Galacticus} \citep{BENSON2012175} to predict the distribution of halos that can host a galaxy of a given magnitude and redshift as a function of the astrophysical model. By forward-modeling JWST galaxies detected by the CEERS and NGDEEP surveys \citep{bagley2023generation,leung2023ngdeep,finkelstein2023complete}, we account for the possible halo properties for each observed galaxy, and determine the best-fit values of the physical parameters, capturing their uncertainties and degeneracies. This procedure allows us to robustly assess consistency with the predictions of galaxy formation models, accounting for the significant theoretical uncertainty of high-$z$ galaxy population predictions.

We directly model the likelihood of observing individual galaxies in apparent magnitude--redshift  ($m_{\mathrm{UV}}$--$z$) space, accounting for the limited volume and depth of specific surveys, and for photometric redshift uncertainties. Thus, our framework does not rely on derived quantities such as the stellar mass or summary statistics like the UVLF. Moreover, we fully incorporate selection effects and statistical uncertainties in the data as well as observational uncertainties on measured quantities (e.g., on photometric redshifts). Our approach is modular and can be applied to any underlying galaxy formation model(s) and high-$z$ galaxy population observations of interest.

This paper is organized as follows. In Section~\ref{sec:model}, we describe our forward model of high-$z$ galaxy populations. In Section~\ref{sec:like}, we present our inference framework, summarize the JWST data used in our fit, and describe our posterior sampling. In Section~\ref{sec:results}, we present our results, focusing on astrophysical parameter constraints, the UVLF, and inferred halo mass distribution and scaling relations for high-$z$ JWST galaxies. We discuss our results in Section~\ref{sec:discussion}; we conclude in Section~\ref{sec:conclusions}. We assume a fiducial cosmology with $H_{0}=70\, \mathrm{km~s^{-1}~Mpc^{-1}}, \Omega_{m}=0.286, \Omega_{\Lambda}=0.714, \Omega_{b}=0.047, \sigma_{8}=0.82, n_{s}=0.9665,$ and $N_{\mathrm{eff}}=3.046$~\citep{HinshawWMAP}.

\section{Forward Model}
\label{sec:model}

\begin{figure}
    \centering
    \includegraphics[width=0.75\linewidth]{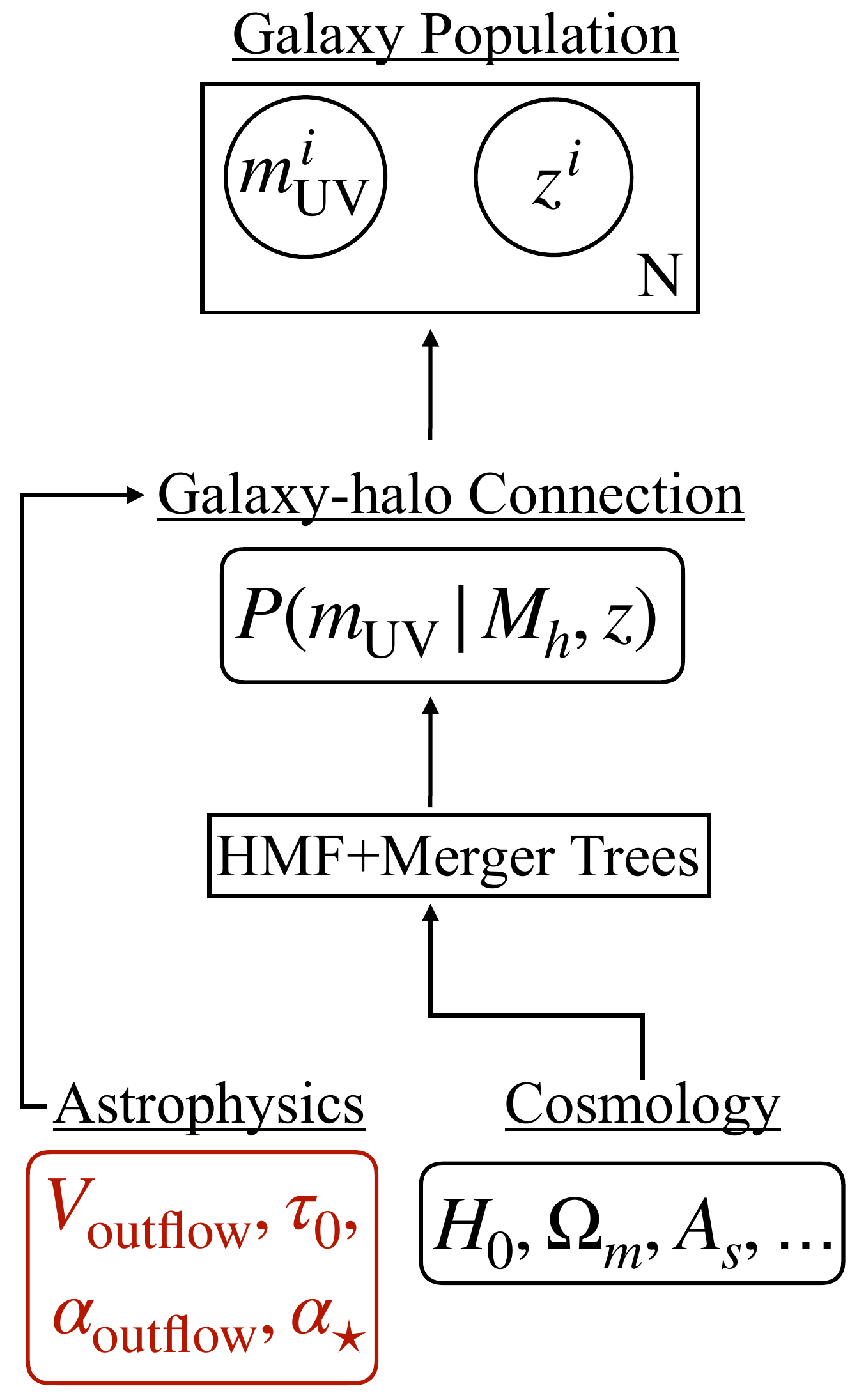}
    \caption{Outline of the forward-modeling pipeline presented in this work. N stands for the total number of high-$z$ galaxies, where the individual apparent magnitudes are $m_\textrm{UV}^i$ and redshifts $z^i$, $i\in\{1,N\}$. The red highlighted parameters are varied in the fit to JWST data. We define the cosmological parameters in Section \ref{sec:intro}, the halo mass function model and merger tree algorithm are defined in Section \ref{ss:mergertrees}, the astrophysical model is defined in Section \ref{ss:galaxyevo}, the model of galaxy evolution is defined in Section \ref{ss:galaxyevo}, and the model of the connection between galaxy and halo properties is given in Section \ref{sec:like}, and the full population-synthesis prediction is outlined in Section \ref{sec:like} and \ref{sec:uvlf}.}
    \label{fig:pipeline}
\end{figure}

We model the population of high-$z$ galaxies observed by JWST using the semi-analytic galaxy formation model \textsc{Galacticus}.\footnote{\url{https://github.com/galacticusorg/galacticus}} Specifically, we calculate halo merger trees and mass functions using extended Press--Schechter (ePS) theory. We apply a simple, semi-analytic galaxy evolution model to our merger trees to predict the galaxy--halo connection as a function of redshift. We convolve this prediction with the halo mass function to predict galaxy number density as a function of apparent magnitude and redshift. We summarize our model in Figure~\ref{fig:pipeline}. The following subsections describe each component of our model.

\subsection{Dark Matter Halo Evolution}\label{ss:mergertrees}

To model the population and evolution of halos, we use the ePS formalism to predict halo abundances and merger histories. Specifically, the halo mass function (HMF) is determined from an excursion set formalism using a remapped barrier from \cite{Sheth2001}, calibrated for ellipsoidal collapse of overdensities. Separately, we construct merger trees according to the \cite{Cole2000} algorithm, with a maximum branching probability in a single time step of 0.1, and a maximum fractional change of 0.1 in mass due to sub-resolution accretion. We calculate branching probabilities according to the \cite{Parkinson2008} modifier function, with parameters $G_{0}=0.57$, $\gamma_{1}=0.38$, and $\gamma_{2}=-0.01$. We build merger trees out to a maximum redshift of $z=200$ and a minimum halo mass of $\mathrm{max}(10^{6}, 10^{-3}\times M_{h})$, where $M_{h}$ is the root node mass of each merger tree. We generate 2000 merger trees per decade in halo mass with final masses spanning $10^{8}-10^{11.5} M_{\odot}$ anchored at three different redshifts---$z=8,~ 12$, and $16$---to generate predictions for the underlying population of dark matter halos that can host the high-$z$ galaxies observed by JWST. We only output \textsc{Galacticus} data at these three redshifts for computational efficiency, and we describe our method for interpolating between these redshifts below.  

\cite{Yung2024} show that halo mass function fitting functions, such as that of \cite{Sheth2001} used here, can become inaccurate at high redshifts compared to recent, high resolution N-body simulations. As we show in \S\ref{sec:haloMassDistribution}, our model predicts that high-$z$ galaxies observed by CEERS and NGDEEP occupy halos with typical masses below $\log_{10}(M_\mathrm{h}/\mathrm{M}_\odot)=10.5$. \cite{Yung2024} show that, for these halo masses, the \cite{Sheth2001} mass function is accurate to better than $\approx 15\%$ at $z=10$, and to better than $\approx 50\%$ at $z=15$. Given the current precision of high-$z$ galaxy population observations and the fact that most observed galaxies are concentrated at the lower end of the redshift range we consider, the HMF and merger tree methods we adopt are sufficiently accurate. However, improved halo models will be important for precision inference from future datasets with larger high-$z$ galaxy samples.

\subsection{Galaxy Evolution}
\label{ss:galaxyevo}

Based on the halo merger trees, we model the evolution of the corresponding galaxy population using the semi-analytic model \textsc{Galacticus} \citep{BENSON2012175}. \textsc{Galacticus} defines components for relevant galactic quantities such as the dark matter halo, galactic disk, circumgalactic medium (CGM), etc.\ and evolves each component according to differential equations that describe the physical processes governing its evolution. Variations in the underlying merger history for each halo therefore lead to a diverse population of galaxies.

The basic astrophysical model we use is similar to that described in \citeauthor{Knebe2018}~(\citeyear{Knebe2018}; \S2.2), with default parameter values chosen to approximately match a set of low-$z$ galaxy population measurements, including the galaxy stellar mass function \citep{li_distribution_2009}, K \citep{cole_2df_2001} and b-band \citep{norberg_2df_2002} luminosity functions, the Tully-Fisher relation \citep{pizagno_tully-fisher_2007}, the color-magnitude relation \citep{weinmann_properties_2006}, the distribution of disk sizes \citep{de_jong_local_2000}, the black hole--bulge mass relation \citep{haring_black_2004}, and the star formation history of the Universe \citep{hopkins_evolution_2004}.

While many physical processes impact the properties of high-$z$ galaxies, we specifically focus on varying the efficiency of star formation and stellar feedback-driven outflows. Enhancing the star formation efficiency boosts galaxies' stellar masses and luminosities, while also increasing the rate at which gas is forced out of galaxies due to stellar winds. Conversely, enhancing outflow strength reduces the amount of gas available for star formation, suppressing the rate at which the stellar masses of galaxies grow. We consider variations only to the star formation rate (SFR) and outflow parameters in the disk component of each \textsc{Galacticus} galaxy, since most of the galaxies in our model predictions are disk-dominated. The spheroid component for each model galaxy is evolved similarly, but its parameters are fixed to values that replicate the population of galaxies at low redshifts. We do not model dust, and we assume a standard \cite{Chabrier2001} IMF. We do not explicitly model UV variability due to bursty star formation, beyond what is captured by our merger trees; we discuss these choices in Section~\ref{sec:discussion}.

\begin{figure*}
     \begin{subfigure}[t]{0.49\textwidth}
         \centering
         \includegraphics[width=\linewidth]{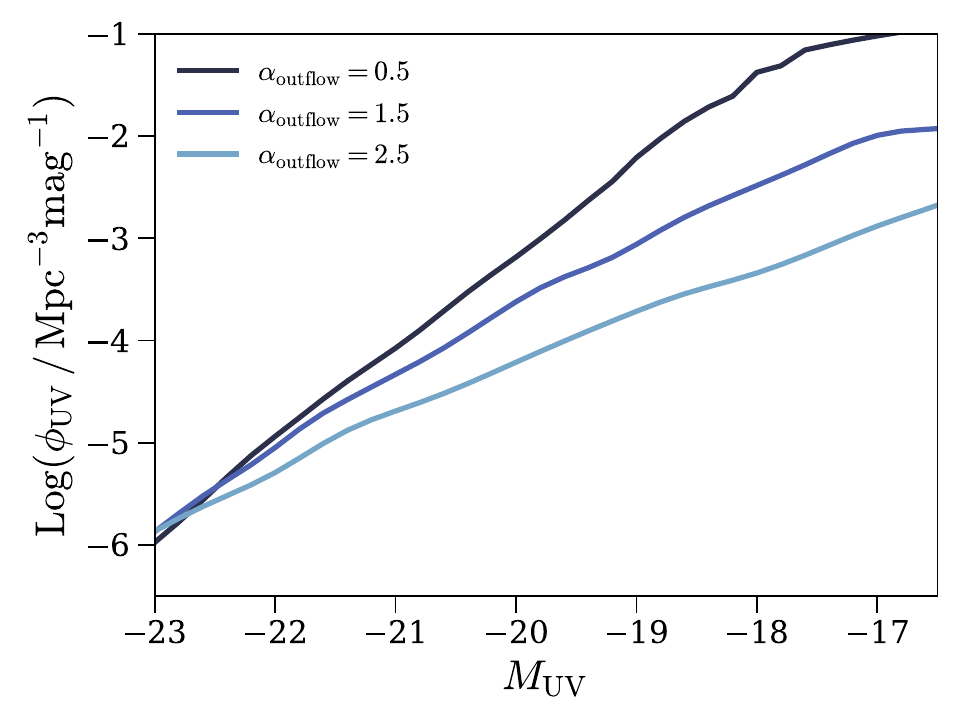}
     \end{subfigure}
     \hfill
     \begin{subfigure}[t]{0.49\textwidth}
         \centering
         \includegraphics[width=\linewidth]{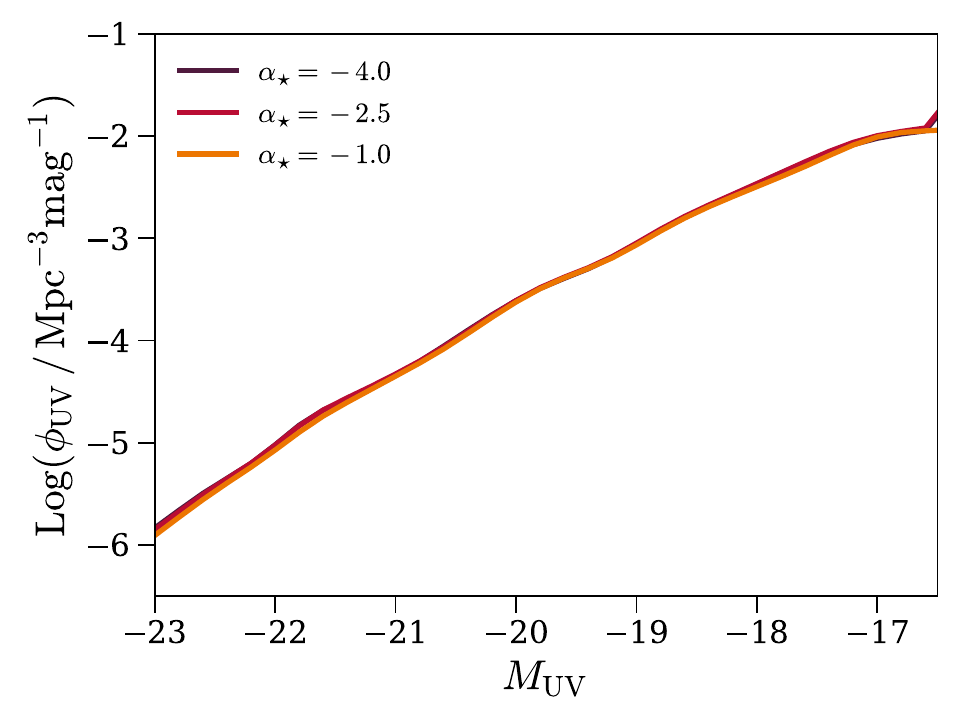}
     \end{subfigure}
     
     \medskip
     
     \begin{subfigure}[t]{0.49\textwidth}
         \centering
         \includegraphics[width=\linewidth]{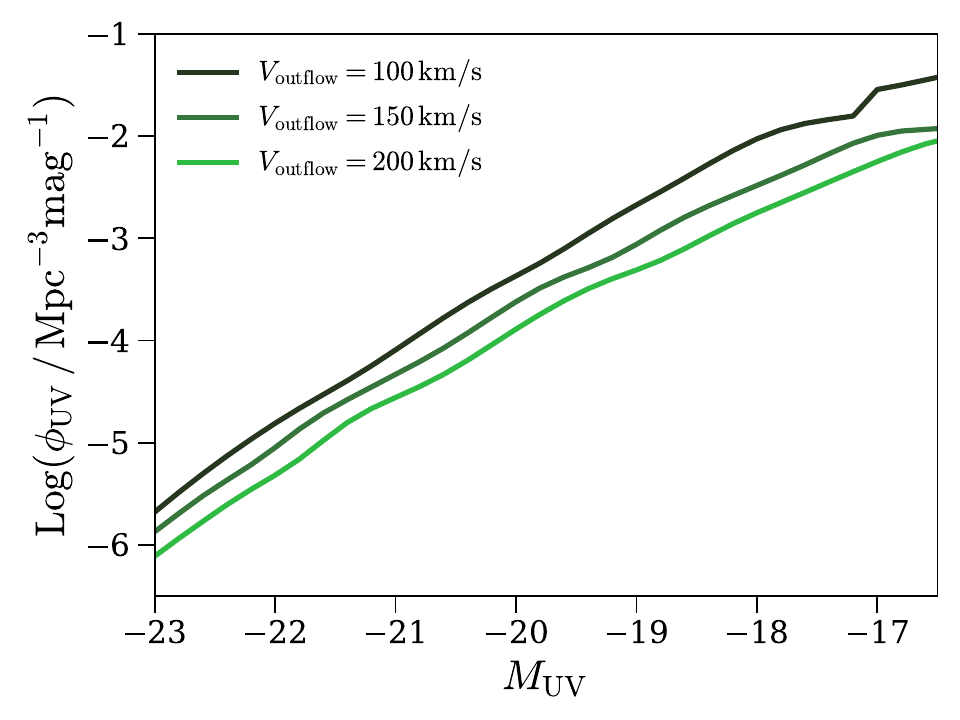}
     \end{subfigure}
     \hfill 
     \begin{subfigure}[t]{0.49\textwidth}
         \centering
         \includegraphics[width=\linewidth]{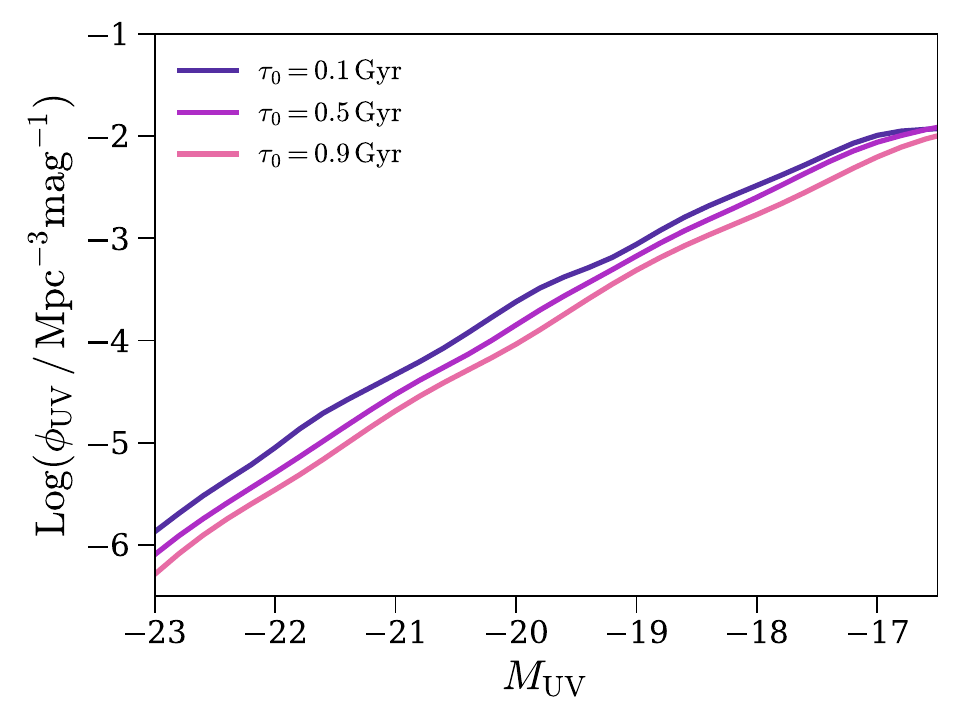}
     \end{subfigure}
     \caption{UV luminosity functions (UVLFs) predicted by our model, at $z=8$. Each panel illustrates the impact of a single astrophysical parameter, while holding the rest of the parameters fixed to the best-fit values obtained in our likelihood analysis. The parameter that is varied is labeled in the legend of each panel. Note that the power-law index parameters that control the halo mass dependence of the star formation rate (top left) and the disk-velocity dependence of the outflow rate (top right) both modifying the tilt of the UVLF, while the parameters controlling the overall rate of the two processes (bottom row of panels) only change the overall normalization of the UVLF.}
\label{fig:astro_params}
\end{figure*}

We model the evolution of gas in the disk component of each galaxy using the differential equation
\begin{equation}
\begin{split}
    \Mdot_{\mathrm{gas}} &= \Mdot_{\mathrm{inflow}} - \Mdot_{\mathrm{outflow}} - \dot{M}_{\star} - \frac{M_{\mathrm{gas}}}{\tau_{\mathrm{bar}}}
\end{split}
\end{equation}
where the overdot denotes a derivative with respect to physical time, $\Mdot_{\mathrm{inflow}}$ is the inflow rate, $\Mdot_{\mathrm{outflow}}$ is the outflow rate due to feedback, $\dot{M}_{\star}$ is the rate at which gas is turned into stars, and $\tau_{\mathrm{bar}}$ is the bar instability timescale calculated following \cite{Efstathiou1982}.
To determine the rate at which cold gas accretes onto the galaxy due to cooling, first the cooling function is computed using the \textsc{CLOUDY} code under the assumption of collisional ionization equilibrium with no
molecular contribution \citep{2023RMxAA..59..327C}. The cooling function is used to determine the gas cooling timescale as a function of radius given by 
\begin{equation}
   t_{\mathrm{cool}} (r) = \frac{3}{2} \frac{k_{\mathrm{B}}T(r) n_{\mathrm{tot}}(r)}{\Lambda}, 
\end{equation}
where $k_{\mathrm{B}}$ is the Boltzmann constant, $T(r)$ is temperature of the gas, $n_{\mathrm{tot}}(r)$ is the total particle number density, and $\Lambda$ is the cooling function. Next, we determine the radius, $r_\mathrm{cool}$ in the halo within which gas has been able to cool by the present time by solving $t_\mathrm{cool}(r_\mathrm{cool}) = t_\mathrm{now}-t_\mathrm{start}$, where $t_\mathrm{now}$ is the present time for the halo, and $t_\mathrm{start}$ is the time at which cooling began in the halo. \cite{White1991} simply set $t_\mathrm{start}=0$, such that $t_\mathrm{cool}(r_\mathrm{cool}) = t_\mathrm{now}$---we make the same choice and solve this equation for $r_\mathrm{cool}$.

The radius from which gas is able to infall and accrete onto the galaxy is determined from the smaller of the cooling radius and the freefall radius, and is used as input to the gas accretion rate which we calculate according to \cite{White1991}: $\dot{M} = 4 \pi \rho(r_\mathrm{cool}) r^2_\mathrm{cool} \dot{r}_\mathrm{cool}$, where $\rho(r)$ is the density of gas in the circumgalactic medium. The accretion rate is then combined with a model for the cold-mode inflow and angular momentum content of the galaxy to determine the rate $\dot{M}_{\mathrm{inflow}}$ at which circumgalactic gas flows into the galaxy.

To model stellar feedback, we parameterize the rate at which the disk loses gas due to outflows $\Mdot_{\mathrm{outflow}}$ as a power law of disk velocity, 
\begin{equation}
 \Mdot_{\mathrm{outflow}}=\left(\frac{\Vout}{V_{\mathrm{disk}}}\right)^{\alphaout}\Psi, 
\end{equation}
where $V_{\mathrm{disk}}$ is the circular velocity of the disk at the exponential profile scale radius and $\Vout$ is the characteristic velocity scale at which the supernova (SNe)-driven outflow rate equals the SFR of the disk component (corresponding to an outflow mass-loading factor of $\sim 1$), $\alphaout$ is the velocity power-law index for the outflow rate, and $\Psi$ is the instantaneous SFR defined below. We consider variations in both $\Vout$ and $\alphaout$ to test how the velocity dependence and overall outflow rate affect the observed population of galaxies.

The total stellar mass of the disk evolves according to 
\begin{equation}
    \Mdot_{\star} = \Psi - \dot{R} - \frac{M_{\star}}{\tau_{\mathrm{bar}}} - \frac{M_{\mathrm{gas}}}{M_{\mathrm{gas}}+M_{\star}}\Mdot_{\mathrm{tidal}},
\end{equation}
where $\dot{R}$ is the rate of mass recycling from stars. For the recycling rate, we assume instantaneous recycling $\dot{R}=R\Psi$ with a recycled fraction of $R=0.46$, which is typical for a stellar population with a \cite{Chabrier2001} IMF and an age of 10~Gyr, as described in \cite{2010PhR...495...33B}. 
The instantaneous SFR is defined as
\begin{equation}
    \Psi = \frac{M_{\mathrm{gas}}}{\tau_{\star}},
\end{equation} 
where $\tau_{\star}$ is the instantaneous star formation timescale, modeled as a power law in the velocity of the disk,
\begin{equation}
    \tau_{\star} = \tau_{0}\left(V_{\mathrm{disk}}/50\mathrm{km\ s}^{-1}\right)^{\alphastar},
\end{equation}
with a power law index $\alphastar$, and $\tau_{0}$ is the normalization of star formation timescale.

Finally, the stellar luminosity is computed from the SFR as
\begin{equation}
    \dot{L}_{\mathrm{\star}} = \Psi \mathcal{L}_{\lambda}(t_{0}-t,Z_{\mathrm{gas}})-\frac{L_{\mathrm{\star}}}{\tau_{\mathrm{bar}}}-\frac{L_{\mathrm{\star}}}{M_{\mathrm{gas}}+M_{\mathrm{\star}}}\Mdot_{\mathrm{tidal}},
\end{equation}
where $\mathcal{L}_{\lambda}(t,Z_{\mathrm{gas}})$ is the luminosity-to-mass ratio in a given band for a stellar population of age $t$ and metallicity of the gas, $Z_{\mathrm{gas}}$. The luminosity-to-mass ratio per band is obtained using the \textsc{FSPS} package \citep{Conroy2009}, again assuming a \cite{Chabrier2001} IMF. Specifically, \textsc{FSPS} is used to find the spectrum of a simple stellar population with masses following given by the IMF and total mass normalized to $1\mathrm{M}_\odot$ as a function of that population's age and metallicity. Then, at each age and metallicity, the spectrum (after being shifted to the redshift of interest) is multiplied by the appropriate filter throughput curve and integrated to find the total in-band luminosity.\footnote{For the F227W filter that we focus on in this work, the total system throughput curve was taken from the \href{https://jwst-docs.stsci.edu/jwst-near-infrared-camera/nircam-instrumentation/nircam-filters\#NIRCamFilters-Filterlists}{JWST User Documentation}.}

\subsection{UVLF Reconstruction}
\label{sec:uvlf_reconstruction}

\begin{figure}
    \centering
    \includegraphics[width=\linewidth]{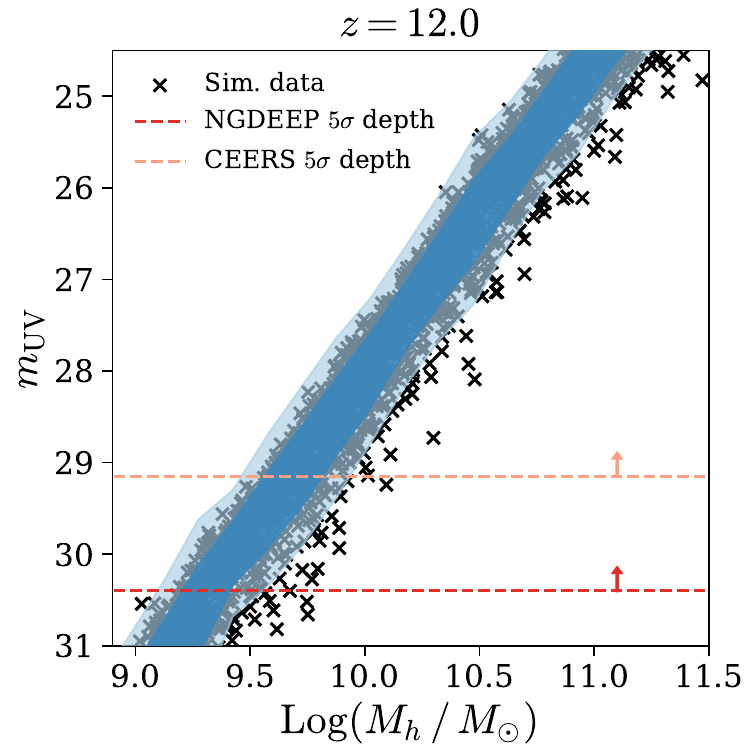}
    \caption{Apparent magnitude--halo mass distribution of predicted galaxies at $z=12$ (black crosses) and the corresponding reconstructed distribution (blue bands), where shaded regions show $1$ and $2\sigma$ contours. Dashed lines indicate the $5\sigma$ limiting depths of the NGDEEP (red) and CEERS (orange) surveys. The blue-shaded reconstructed distribution is used in our likelihood analysis in Section \ref{sec:like}.}
    \label{fig:fit}
\end{figure}

Each \textsc{Galacticus} realization produces a discrete set of halo masses with corresponding magnitudes and redshifts. To reconstruct the corresponding $m_\text{UV}$--$z$ distribution, we assume that, for a given halo mass and redshift, the magnitude is drawn from a Gaussian 
\begin{equation}
    m_{\mathrm{UV}}(M_{h},z) \sim \mathcal{N}(\mu(M_{h},z),\sigma(M_{h},z)).
\end{equation}
We then fit the mean and variance of the Gaussian distribution to the realizations obtained from the semi-analytic model, in bins of width $\Delta\,\log\,M_{h}=0.15$ at each redshift for which we generate data ($z=8$, $12$, and $16$). As halo mass decreases at a given redshift, the scatter in our predicted magnitude--halo mass relation grows, and the resulting distribution becomes less Gaussian. To prevent artificial up-scatter of faint galaxies into brighter magnitude bins due to this modeling choice, we truncate the Gaussian distributions at 10\% of the brightest magnitude in a given mass bin. The result of this fitting procedure is illustrated in Figure~\ref{fig:fit} for our best fit parameter value combination. While our fit misses some of the down-scatter present in the simulated galaxy population due to the non-Gaussianity of the underlying data, this does not bias our predicted UVLFs at a level that would impact our inference.

Finally, to reconstruct UVLFs, we linearly interpolate the resulting $M_{\mathrm{UV}}-M_{h}$ relation for intermediate redshift bins with width $\Delta\,z=0.5$ to model redshift evolution. To validate this procedure, we compare our $z=12$ model realizations to the prediction from interpolating between our $z=8$ and $z=16$ realizations; we find that differences in the resulting distributions are negligible. The UVLF can then be calculated by marginalizing over halo mass weighted by the abundance of each halo
\begin{equation}
    \phi_{\mathrm{UV}}(\muv,z) = \int \mathrm{d}M_{h}\, P(\muv|M_{h},z)\frac{\mathrm{d}n}{\mathrm{d}M_{h}},\label{eq:phi_uv}
\end{equation}
where the halo mass function $\frac{\mathrm{d}n}{\mathrm{d}M_{h}}$ matches the one used to generate the galaxy merger trees in Section~\ref{ss:mergertrees}. Note that, although we predict a distribution of galaxies for each underlying astrophysical model, the theoretical UVLF itself is not stochastic.

\subsection{Model Illustration}

In our likelihood analysis (Section~\ref{sec:like}), we will vary $\Vout$, $\alphaout$, $\tau_{0}$, and $\alphastar$ as free parameters; Figure~\ref{fig:astro_params} illustrates how they impact the observed UVLF, computed using the procedure described in the above Section. When varying the power-law exponents $\alphastar$ and $\alphaout$ for both feedback and SFR, the UVLF slope changes as the relative contributions of halos of different masses to each luminosity bin change. In contrast, modifying the normalization parameters $\tau_{0}$ and $\Vout$ alters the overall normalization of the UVLF, as the gas and stellar mass of every halo are affected by these parameters.

Previous work has hinted that a redshift evolution of the SFR model parameters may be necessary to reconcile low and high-$z$ UVLFs (e.g., \citealt{Hou2016,Yung230404348}). While our SFR changes with redshift due to halo growth captured in the merger trees, we do not explicitly model the evolution of any astrophysical parameters. After determining allowed ranges of our model parameters using high-$z$ JWST data, we test whether our model accurately predicts lower-$z$ UVLFs and stellar mass functions. This allows us to test whether the simplified galaxy-evolution model can simultaneously describe the evolution of galaxies over a wide redshift range.

\section{Analysis}
\label{sec:like}

\subsection{Likelihood}

Consider an observed set of galaxy candidates with apparent magnitudes and redshifts, denoted by $\mathcal{D}\equiv (m_{i},z_{i})$. We model the likelihood of observing this data as an inhomogeneous Poisson process, with the occurrence rate for a given magnitude and redshift pair given by the UVLF. We assume that, if galaxy candidate's apparent magnitude and redshift were known exactly, the likelihood can be written as
\begin{equation}\label{eq:like}
    \begin{split}
    \mathcal{L}(\mathcal{D}|\theta) = 
    &\prod_{s}\Bigg\{ \\
        & \mathrm{exp}\left[- \int \mathrm{d}z\int\mathrm{d}\muv \frac{\mathrm{d}V^{s}_{\mathrm{eff}}(\muv,z)}{\mathrm{d}z}\phi_{\mathrm{UV}}(m_{\mathrm{UV}},z;\theta)\right] \\
        &\prod_{i}^{N_{\mathrm{gal,s}}}\phi_{\mathrm{UV}}(m_{\mathrm{UV},i},z_{i};\theta)\frac{\mathrm{d}V^{s}_{\mathrm{eff}}(m_{\mathrm{UV},i},z_{i})}{\mathrm{d}z}\Bigg\} ,
    \end{split}
\end{equation}
where $\theta=\{\Vout,\alphaout,\tau_0,\alphastar\}$ is the vector of free parameters; the first product is over a collection of different surveys $s$ where $V_{\mathrm{eff}}^{s}$ is the effective survey volume; $\phi_{\mathrm{UV}}(m_{\mathrm{UV}},z;\theta)$ is the predicted UVLF, evaluated at apparent magnitude $m_{\mathrm{UV}}$ and redshift $z$, and calculated for a given parameter vector $\theta$; and the second product is over all observed galaxy-candidate objects in survey $s$, $i\in\{1,N_\textrm{gal, s}\}$. We assume that different surveys are uncorrelated, and we evaluate the likelihood over a grid of $\theta$ parameters to infer the parameter values consistent with CEERS and NGDEEP data.

We go beyond the likelihood above by accounting for photometric redshift uncertainties. Specifically, we marginalize over the redshift distribution for each observed galaxy candidate by replacing 
\begin{align}
        &\phi_{\mathrm{UV}}(m_{\mathrm{UV},i},z_{i})\frac{\mathrm{d}V_{\mathrm{eff}}(z_{i})}{\mathrm{d}z}&\nonumber \\ &\rightarrow
        \int_{z_{\mathrm{min}}}^{z_{\mathrm{max}}} \mathrm{d}z\,P(z)\phi_{\mathrm{UV}}(\muv,z)\frac{\mathrm{d}V_{\mathrm{eff}}(\muv,z)}{\mathrm{d}z},&
        \label{eq:pz}
\end{align}
where $P(z)$ is the probability that a given galaxy candidate has a true redshift $z$, and $\phi_\text{UV}(m_\text{UV}, z)$ is given by Equation \ref{eq:phi_uv}. 

We model redshift uncertainties as follows. First, we note that the redshift uncertainty varies significantly between the galaxy candidates we consider, ranging from $\sigma_{z} \approx 0.01$ to $2.0$. To mimic the color cuts used to select high-$z$ galaxy candidates for each survey catalog, we set the lower bound on the redshift integral in Equation~\ref{eq:pz} to $z_{\mathrm{min}}=8.5$, and we set the upper bound to $z_{\mathrm{max}}=16$; this range approximately covers the full span of possible redshifts for all candidates we consider~\citep{leung2023ngdeep,finkelstein2023complete}. We then normalize $P(z)$ to unity in the range $z_{\mathrm{min}}$ to $z_{\mathrm{max}}$. For reference, among the CEERS and NGDEEP galaxy candidates we consider, the objects with the highest probability to be $z>16$ have $P(z>16) \approx 0.05$. On the other hand, the contribution from redshifts near $z_{\mathrm{min}}$ can be significant for galaxy candidates with median redshifts close to the CEERS and NGDEEP selection criteria, $z_{\mathrm{median}}>8.5$. Instead of modeling the color cuts that each survey uses to select high-$z$ galaxies, we truncate the integral of the predicted probability distribution at $z_\text{min}=8.5$ and renormalize the resulting distribution. Apparent magnitudes uncertainties, which are typically in the range $\sigma_{\muv}\sim 0.1$ are negligible for our analysis.

\begin{table*}
    \centering
    \begin{tabular}{|c|c|c|c|c|c|c|c|}
        \hline
        & Parameter & Physical Interpretation & Prior & $N_{\mathrm{samples}}$ & Best fit & 68\% CI & 95\% CI \\
        \hline
        \multirow{2}{*}{Outflow} &  $V_{\mathrm{outflow}}\ [\mathrm{km/ s}]$ & Outflow rate normalization & $\mathcal{U}(50, 500)$ & 46 & $150$  & $[110,240]$  & $[90,430]$ \\
        & $\alphaout$ & Outflow rate power-law index & $\mathcal{U}(0.125 , 2.5)$ & 20 & 1.5 & $[0.8,1.6]$ & $[0.6,2.0]$ \\
        \hline
        \multirow{2}{*}{SFE} & $\tau_{0}~[\mathrm{Gyr}]$ & Star formation timescale & $\mathcal{U}(0.1, 1.0)$ & 10 & 0.1 & $[0.1,0.2]$ & $[0.1,0.5]$ \\
        & $\alphastar$ & Disk velocity power law index & $\mathcal{U}(-4.0, -0.5 )$ & 8 & $-4.0$ & $[-4.0,-2.1]$ & $[-4.0,-0.8]$ \\
        \hline
    \end{tabular}
    \caption{Summary of the astrophysical parameters we vary along with their prior ranges, the number of linearly-spaced samples in that range, best-fit values, and the 68\% and 95\% confidence intervals determined by our likelihood analysis. $\mathcal{U}(a,b)$ denotes a uniform distribution between $a$ and $b$.}
    \label{tab:astro_params}
\end{table*}

\subsection{Data}
\label{sec:data}

We compare our predictions to observed galaxies from the NGDEEP \citep{bagley2023generation,leung2023ngdeep} and CEERS \citep{finkelstein2023complete} surveys conducted using JWST NIRCam. These surveys are complementary in their coverage of the bright (CEERS) and faint (NGDEEP) ends of the high-$z$ galaxy population. We use the reported estimates of the effective volume for each survey (in magnitude and redshift bins), and assume that neither field is significantly lensed. We only use the observed luminosities in the F277W filter, which is close to the rest-frame UV magnitude of $\sim 1500 \r{A}$ for the redshifts of interest, for which all observed galaxy candidates have reported apparent magnitudes. The NGDEEP survey targets the Hubble Ultra-Deep Field (HUDF)-Par2 field with a survey area of $\sim 5\, \mathrm{arcmin^{2}}$, achieving a $5\sigma$ detection limit of $30.4$ mag in the F277W band. The CEERS survey targets the CANDELS Extended Groth
Strip (EGS) field, covering a total area of $88.1\, \mathrm{arcmin^{2}}$ and reaching a $5\sigma$ detection limit of $29.15$ mag in the F277W band.

\begin{figure*}
    \centering     \includegraphics[width=\linewidth]{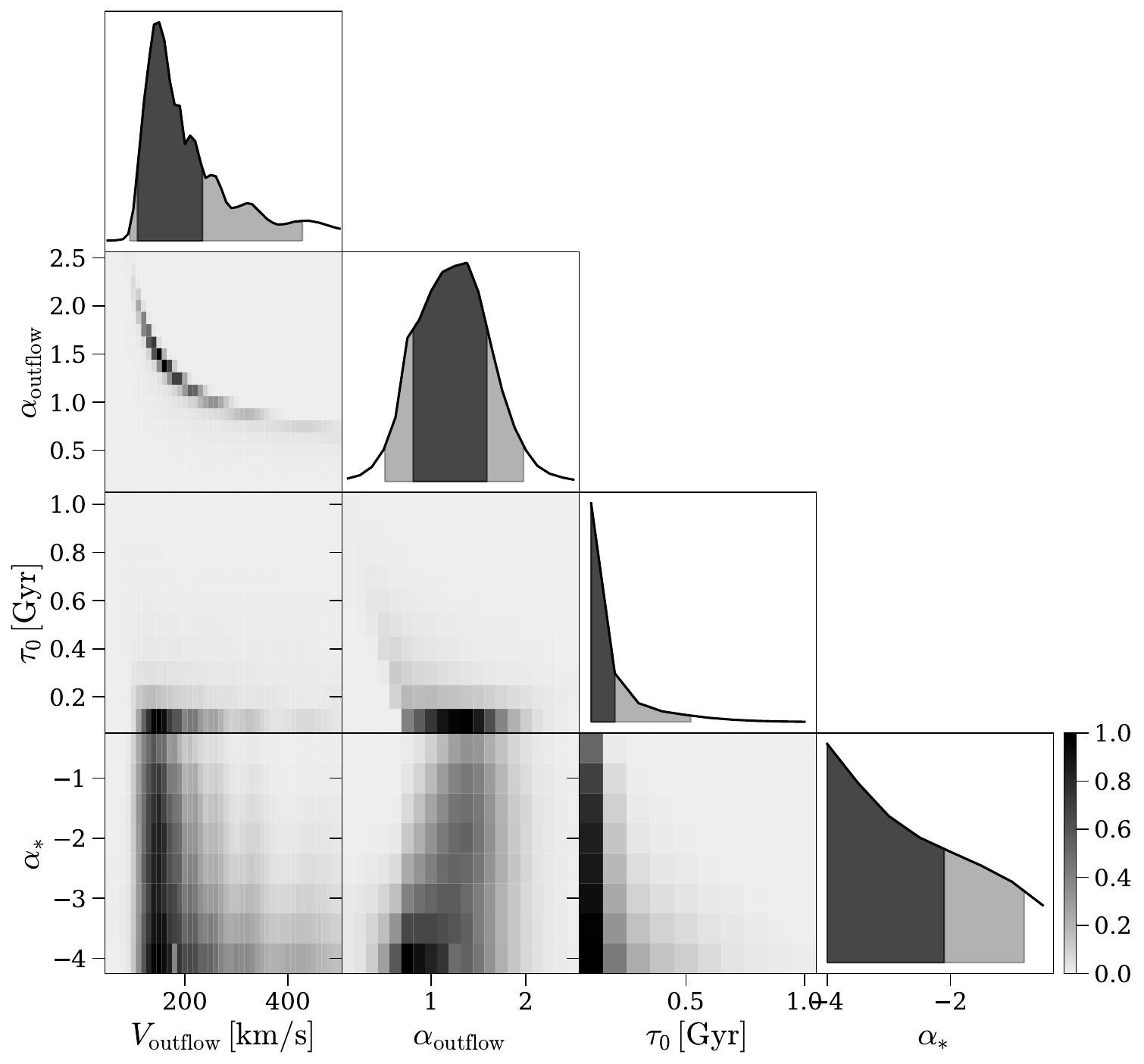}
    \caption{Posterior probability distribution of the astrophysical parameters, obtained from our likelihood analysis. The color scale shows the likelihood ratio relative to the best-fit value. In the one-dimensional marginalized posteriors, dark and light shaded regions denote $68\%$ and $95\%$ confidence intervals, respectively. The corresponding prior ranges, number of samples, best-fit values, and confidence intervals are summarized in Table~\ref{tab:astro_params}. }
    \label{fig:triangle}
\end{figure*}

We analyze a total of 33 galaxy candidates identified in \cite{leung2023ngdeep} (Table 4), and 86 galaxy candidates from \cite{finkelstein2023complete} (listed in Tables 3, 6, and 7); this sample probes down to the $5\sigma$ magnitude depth of each survey. We assume negligible uncertainty in the magnitude, and we account for the two-sided Gaussian uncertainty on the measured photometric redshift, with the mean and $1\sigma$ values reported in each study. Finally, we reconstruct the completeness function for both surveys using the effective volumes quoted in Table 2 of \cite{leung2023ngdeep} and Table 4 of \cite{finkelstein2023complete}. We linearly interpolate these effective volume estimates between magnitude bins, assuming no redshift dependence within each redshift bin.

\subsection{Sampling}
\label{sec:sampling}

We assume uniform prior probabilities on each of the four free parameters of our model, $\theta=\{\Vout,\alphaout,\tau_{*},\alphastar\}$ and take into account the CEERS and NGDEEP data described above in order to evaluate the likelihood given in Equation \ref{eq:like}. We sample the posterior probability distribution at 73,600 parameter combinations, requiring a total runtime of $\sim 10^4$ CPU hours. We show the resulting posterior reconstruction in Figure \ref{fig:triangle}. The global maximum-likelihood values are summarized in Table \ref{tab:astro_params} as the best-fit values. We discuss and interpret our results below.

\section{Results}\label{sec:results}

We next interpret our key result shown in Figure~\ref{fig:triangle}, compare the predicted and observed $m_{\mathrm{UV}}$--$z$ distributions and UVLFs, infer the host halo mass distribution corresponding to the observed galaxies, and compare our predictions to low-redshift galaxy population statistics.


\subsection{Parameter Inference}

\emph{Outflow rate}. The best-fit value for $\Vout$ is $150 ~\mathrm{km\ s}^{-1}$, which corresponds to the disk velocity of a $3\times 10^{10} M_{\odot}$ halo at the redshifts of interest (Appendix~\ref{app:outflow_sfr} shows the outflow rate as a function of halo mass for our best-fit astrophysical parameter combination). The posterior probability distribution for this parameter is skewed toward high values, with $90 < \Vout/\mathrm{km\ s}^{-1} < 420$ at $95\%$ confidence. The best-fit value of the power-law exponent of the outflow rate, $\alphaout$ is $1.5$. The posterior is mildly skewed toward low values of $\alphaout$, with $0.6 < \alphaout < 2.0$ at $95\%$ confidence. The skewness of the marginalized probability distribution for $\alphaout$  comes from a degeneracy with $\Vout$, as seen in Figure~\ref{fig:astro_params}. Along this degeneracy, the amplitude of the UVLF at $-19.5~\mathrm{mag} \lesssim\Muv\lesssim -19.0~\mathrm{mag}$ is roughly fixed by the data, as seen in Figure~\ref{fig:uvlf_spread}. In particular, models with higher $\alphaout$ and lower $\Vout$ enhance the bright end and suppress the faint end of the UVLF, and vice versa. 

Comparing these results with previous literature, the $\Vout$ posterior distribution is consistent with the $z>8$ supernova feedback outflow velocities from \cite{Lu240602672} of $V_{\mathrm{SN}}=180~\mathrm{km\ s}^{-1}$ and $V'_{\mathrm{SN}}=78\mathrm{km\ s}^{-1}$, 
which play a similar role as $\Vout$, for the \textsc{GALFORM} semi-analytic model. Similarly, our $\alphaout$ posterior is consistent with the \textsc{GALFORM} range of outflow velocity exponents, $1.0\lesssim \gamma_{\mathrm{SN}}\lesssim 3.4$. Note that \cite{Hou2016} and \cite{Lu240602672} show that these outflow parameters must evolve in order to match data at lower redshifts than what we consider, with stronger feedback toward lower redshifts. We leave an exploration of such redshift evolution for future work.

\begin{figure*}
    \centering
     \includegraphics[width=\linewidth]{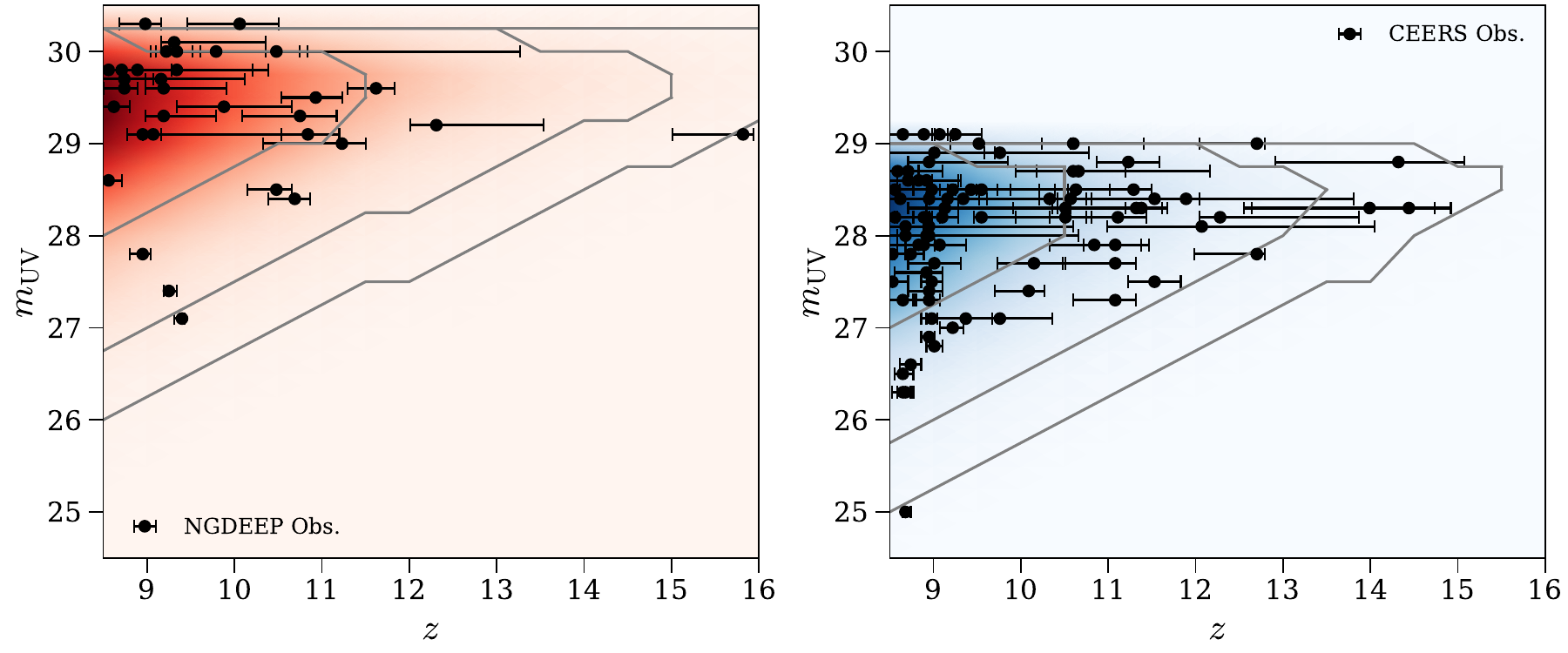}
    \caption{The probability density for observing a galaxy as a function of apparent magnitude and redshift, evaluated at the best-fit parameter values from our analysis (shaded regions). The $1$, $2$, and $3\sigma$ contours are shown as gray lines. We overplot the data points associated with individual galaxies in the CEERS (left) and NGDEEP (right) surveys, including photometric redshift uncertainties (black error bars). The predicted distributions are broadly consistent with the data.}
    \label{fig:muvz_viz}
\end{figure*}

\emph{Star formation rate}. We infer an upper limit on the timescale for star formation of $\tau_{0} < 0.5~\mathrm{Gyr}$ (at $95\%$ confidence), at a disk velocity of $50~\mathrm{km\ s}^{-1}$. This disk velocity is typical of the lowest-mass halos that host CEERS and NGDEEP galaxies in our inference ($\sim 10^9~M_{\mathrm{\odot}}$). Meanwhile, the star formation timescale in $\sim 10^{10}~M_{\mathrm{\odot}}$ halos, which host the majority of the observed galaxies in the sample we consider (see Section~\ref{sec:haloMassDistribution}), can be as short $\sim 6~\mathrm{Myr}$, when evaluated at the best-fit $\alphastar$. We show the star formation timescale as a function of halo mass for our best-fit astrophysical parameter combination in Figure~\ref{fig:moutflow_taustar}. We find that the UVLF is insensitive to star formation timescales below our $95\%$ confidence upper limit on $\tau_0$. This result indicates a preference for effectively instantaneous star formation in our model. The data is not highly sensitive to the halo mass dependence of the star formation efficiency, captured by $\alphastar$.

Our inferred SFRs are higher than values reported in previous literature; e.g., \cite{Baugh2005} reports $\tau_{0}=8$ Gyr and $\alphastar = -3$ by calibrating their semi-analytic model to the abundance of sub-millimeter and Lyman-break galaxies at $0<z<3$. This difference again suggests a redshift evolution of the star formation rate.

\begin{figure*}
    \centering
     \includegraphics[width=\linewidth]{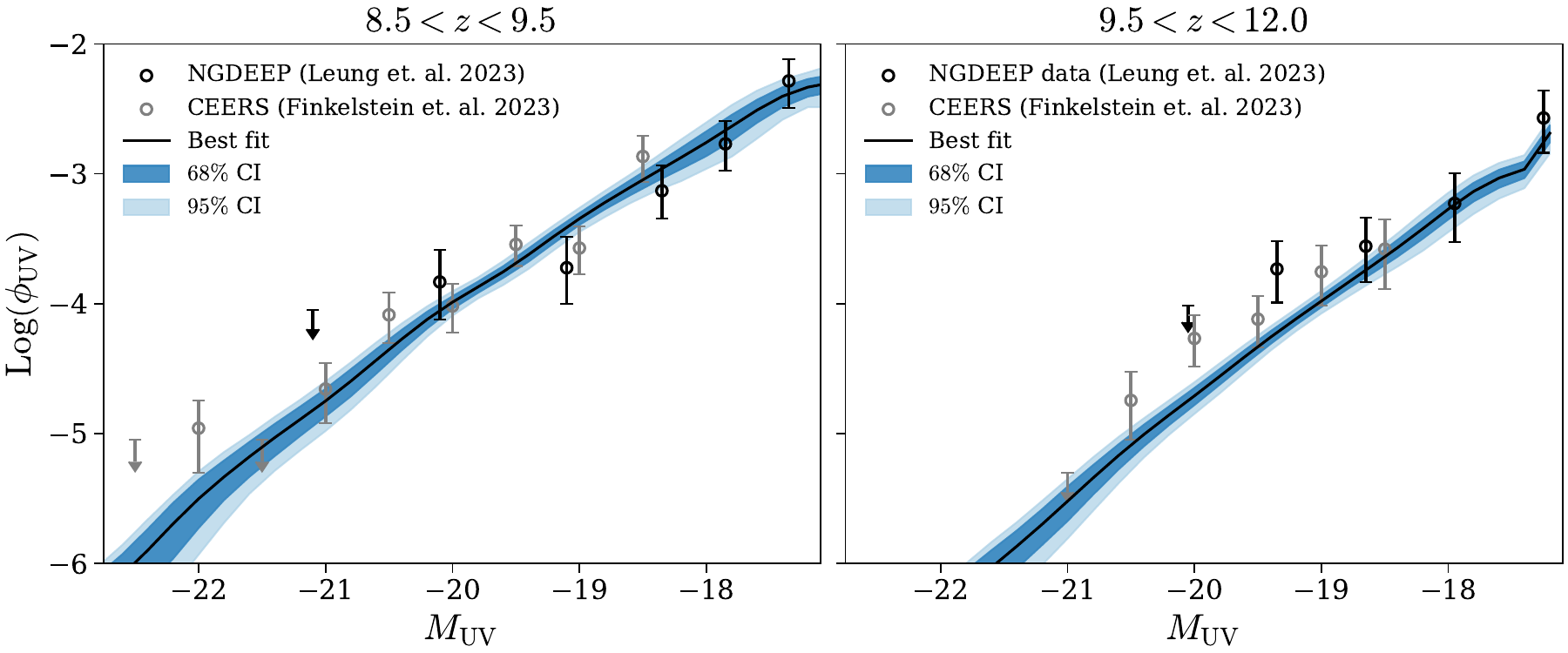}
    \caption{Blue shaded bands show the UV luminosity function (UVLF), in units of $\mathrm{Mpc^{-3} mag^{-1}}$, in redshift bins of $8.5<z<9.5$ (left) and $9.5<z<12.5$ (right). The parameters used to compute the UVLF and its $68\%$ and $95\%$ confidence intervals (shown as dark and light shaded regions, respectively), are computed for the best-fit parameter-value region identified in our likelihood analysis. We overplot binned UVLF estimates and associated error bars from NGDEEP (black; Table 2 of \citealt{leung2023ngdeep}) and CEERS (gray; Table 4 of \citealt{finkelstein2023complete}). Note that our UVLF prediction was not directly fit to these data points, but rather inferred from the underlying Poisson-likelihood analysis in $m_{\mathrm{UV}}$--$z$ space; the UVLF serves as a cross-check of the analysis and indicates agreement with the data. We translate the $5\sigma$ limit on the apparent magnitude from NGDEEP to a redshift limit as a function of absolute magnitude, which causes a slight uptick in the UVLF prediction for the faintest magnitude bins.}
    \label{fig:uvlf_spread}
\end{figure*}

\begin{figure*}
    \centering
     \includegraphics[width=\linewidth]{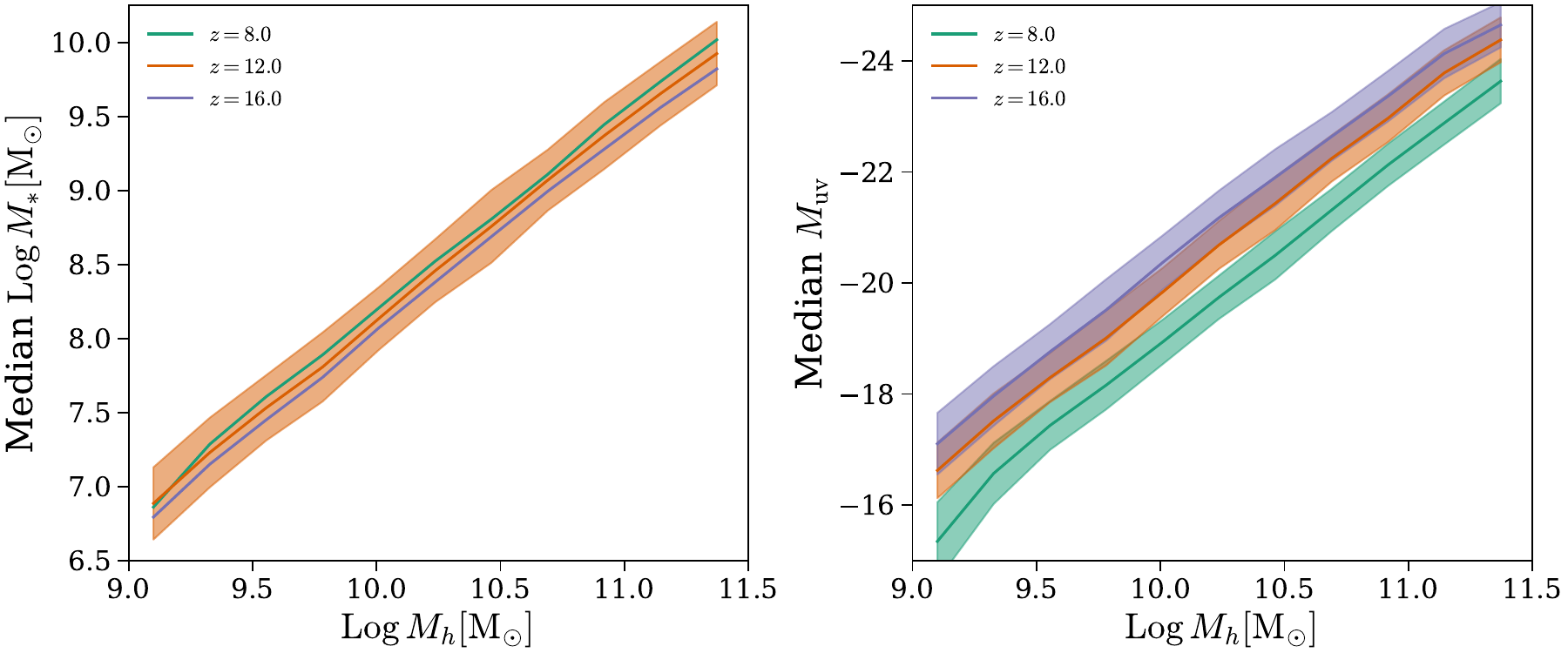}
    \caption{Median (lines) and $1\sigma$ scatter (shaded bands) of stellar mass $M_{\ast}$ (left) and absolute magnitude $\Muv$ (right) as a function of halo mass, for the best-fit parameter values we infer. The scatter in the stellar mass--halo mass relation has little evolution with redshift, so we plot the scatter only for $z=12$.}
    \label{fig:ghc}
\end{figure*}

\subsection{UVLF}\label{sec:uvlf}

Figure~\ref{fig:muvz_viz} shows the $m_{\mathrm{UV}}$--$z$ distribution of high-$z$ galaxies predicted by our best-fit parameters in the \textsc{Galacticus} model, for CEERS and NGDEEP; observed data points are overplotted for each survey. The distribution is a normalized version of the observable that directly enters our likelihood function, given by
\begin{equation}
    P(\muv, z) = \frac{\phi(\muv,z)V_{\mathrm{eff}}(\muv,z)}{\int \mathrm{d}z\,\mathrm{d}\muv\,\phi(\muv,z) V_{\mathrm{eff}}(\muv,z)},
\end{equation}
where $V_\mathrm{eff}$ accounts for for the completeness of each survey. We show the $68\%, 95\%,$ and $99\%$ confidence intervals of the predicted distributions in gray solid lines in the same Figure. We find that roughly $54\%, 88\%,$ and $91\%$ of the data points fall within the respective contours, indicating broad consistency between the data and model. 

By eye, the model slightly underpredicts the abundance of the highest-redshift galaxies in CEERS, so we proceed to test the consistency between the data and the model in more detail with a Peacock test \citep{1983MNRAS.202..615P} (which extends the standard Kolmogorov-Smirnov test to higher dimensions), using the public code \textsc{ndtest}.\footnote{\url{https://github.com/syrte/ndtest}} This yields a $p$-value of $0.05$ and $0.1$ for the NGDEEP and CEERS survey, respectively; the fit to the combined data has a $p$-value of $0.1$, indicating that our predicted distribution is consistent with the data.

In Figure~\ref{fig:uvlf_spread}, we compare the UVLF corresponding to our best-fit parameter values (where uncertainty bands are determined by the $68\%$ and $95\%$ confidence intervals on the parameters) as inferred from our likelihood analysis, to previously-reported reconstruction of the luminosity function from CEERS and NGDEEP \citep{leung2023ngdeep,finkelstein2023complete}. The total reduced chi-squared fit, from all $23$ data points and with four degrees of freedom, is $\chi^{2}_{19}=1.42$. Thus, our predictions fit the binned UVLF reconstructions reasonably well. We emphasize that our likelihood analysis is \textit{not} equivalent to performing a chi-squared minimization on this specific UVLF data; rather, we directly fit the $m_{\mathrm{UV}}$--$z$ distribution using Poisson likelihood. The consistency we observe here validates our approach.

In general, the measurements shown here tightly constrain the UVLF around $-20 \lesssim \Muv \lesssim -19$, with greater allowed spread at fainter and brighter absolute magnitudes. At the very bright end, the UVLF is not well constrained because there are few observed galaxies in this regime, with only one galaxy observed with $\Muv < -21.25$, and five galaxies observed with $\Muv < -20.75$. Meanwhile, at sufficiently faint magnitudes, survey incompleteness dominates the inference. For example, the faintest data points shown in Figure~\ref{fig:uvlf_spread} have a quoted completeness of $\sim 10\%$ in both NGDEEP and CEERS.

\subsection{Stellar mass--halo mass relation}

For the best-fit parameter combination, we plot stellar mass and UV magnitude as a function of halo mass in Figure~\ref{fig:ghc}. Our results are consistent with \cite{Lu240602672}, which calibrates \textsc{GALFORM} to high-$z$ JWST data and low-$z$ galaxy population statistics. We also compare to \cite{Feldmann240702674}, who present a cosmological hydrodynamic simulation designed to study high-$z$ galaxies using the \textsc{FIRE} code; we predict slightly higher stellar masses at fixed halo mass than that study. Our best-fit model predicts a mild evolution in the stellar mass--halo mass relation over the redshift range we consider, with stellar mass slightly increasing over time at fixed halo mass. This result is in contrast with predictions from some empirical models (e.g., \citealt{Behroozi_2019}). Meanwhile, we predict more significant evolution in the $M_{\mathrm{UV}}$--$M_{h}$ relation, with luminosities decreasing over time at fixed halo mass. This finding is similar to \cite{Feldmann240702674}, while \cite{Lu240602672} predicts less significant redshift evolution.

\subsection{Halo mass distribution of high-$z$ galaxies}\label{sec:haloMassDistribution}

\begin{figure}
    \centering
     \includegraphics[width=\linewidth]{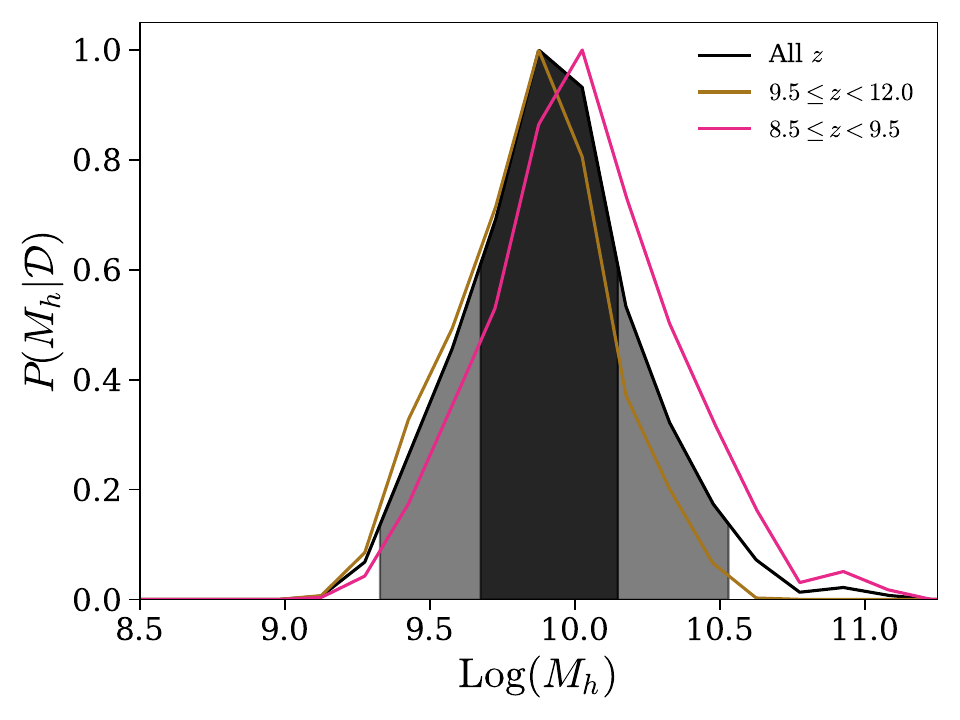}
    \caption{The inferred halo mass distribution of \emph{observed} NGDEEP and CEERS galaxies, evaluated for our best-fit model. We show the distribution for all redshifts (black), where dark (light) shaded regions show $68\%$ ($95\%$) percentiles. We also split the distribution into higher (gold) and lower (magenta) redshift bins.}
    \label{fig:mh_pdf}
\end{figure}

Next, we calculate the best-fit host halo masses of the \emph{observed} galaxies considered in our analysis. The host halo mass distribution follows from our predicted stellar mass--halo mass relation $P(\muv|M_{h},z)$; using Bayes' theorem, we obtain
\begin{equation}
    P(M_{h}|\muv,z) = \frac{P(\muv|M_{h},z)P(M_{h}|z)}{P(\muv)},
\end{equation}
where $P(\muv|M_{h},z)$ is evaluated from \textsc{Galacticus} samples, as described in Section \ref{sec:uvlf_reconstruction}, at the best-fit parameter vector $\theta$; the prior $P(M_{h}|z)$ is determined from the halo mass function; and the evidence $P(\muv)$ is an overall normalization factor. The halo mass distribution for a given collection of observed galaxies is then
\begin{equation}
    P(M_{h}|\mathcal{D}) = \frac{1}{N}\sum_{i}^{N_{\mathrm{gal}}} \int \mathrm{d}z\,P_{i}(z)P(M_{h}|m_{\mathrm{uv,}i},z).
\end{equation}
Figure~\ref{fig:mh_pdf} shows the halo mass distribution averaged over all high-$z$ galaxies observed by CEERS and NGDEEP, split by redshift to show its evolution. We find that the halo mass distribution of observable CEERS and NGDEEP galaxies at $8.5<z<12$ peaks at a virial mass of $10^{10}~M_{\mathrm{\odot}}$, with an uncertainty of $\approx 0.5~\mathrm{dex}$ at $95\%$ confidence; the peak of the halo mass distribution corresponds to $\Muv\approx-19.0$.

\subsection{Comparison to Lower-redshift Data}
\label{sec:lowz}

\begin{figure*}
\begin{subfigure}[c]{0.495\textwidth}
    \centering 
    \includegraphics[width=\linewidth]{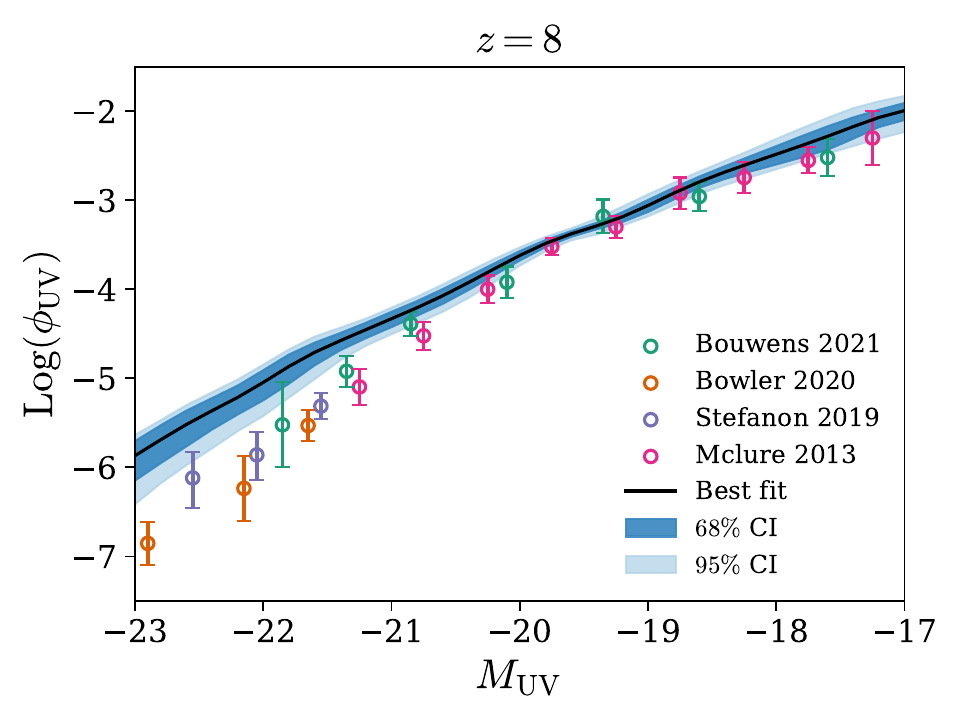}
    \label{fig:hst}
    \centering
\end{subfigure}
\hfill
\begin{subfigure}[c]{0.495\textwidth}    
    \centering\includegraphics[width=.80\linewidth]{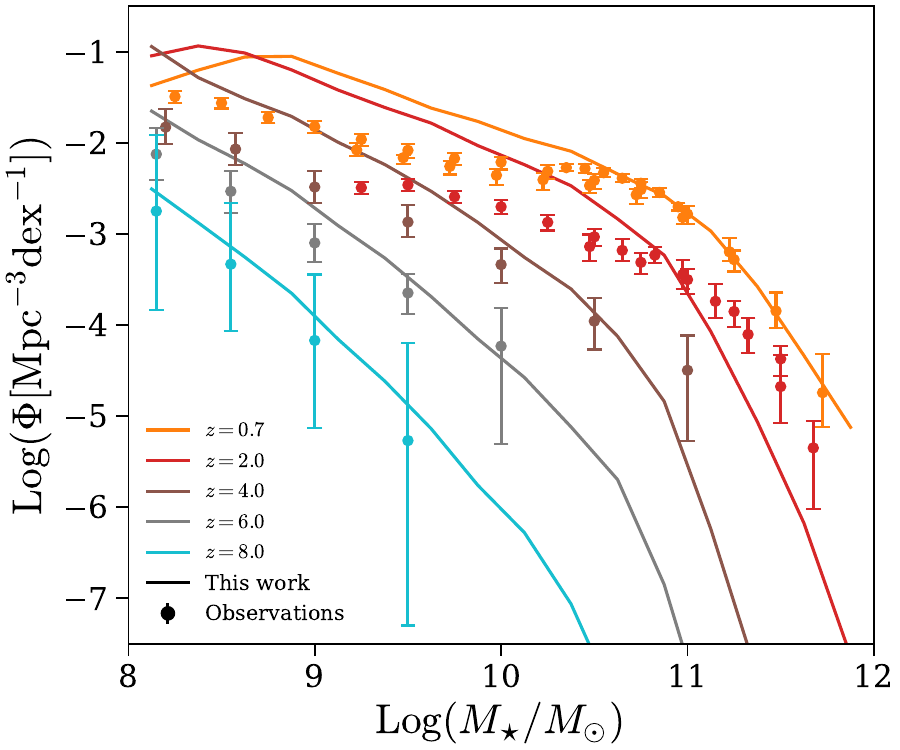}
    \label{fig:z0-comp}
\end{subfigure}
\caption{\textbf{Left}: The predicted UV luminosity function (UVLF) at $z=8$, in units of $\mathrm{Mpc^{-3} mag^{-1}}$, is shown for the best-fit parameter values inferred using JWST data (with $68\%$ and $95\%$ confidence intervals shown by the dark and light blue shaded regions, respectively). The UVLF is computed in the same manner as Figure~\ref{fig:uvlf_spread}, and we overplot HST data from \protect\cite{Bouwens2021, Bowler2020, McLure2013, Stefanon2019} (colored data points with uncertainties). The UVLF that was optimized to fit JWST data at $z\geq 8.5$ fits the HST data well at faint end, and mildly over-predicts the abundance of the brightest galaxies observed by HST at $z=8$. \textbf{Right}: The stellar mass function (SMF) $\Phi$ is shown for $z=0.7-8.0$ for the parameter combination that presents the best fit to JWST data (solid lines). We overplot the observed SMFs from \protect\cite{Behroozi_2019}; specifically, we show data at $4<z<8$ \protect\citep{Song2016}, $2$ \protect\citep{Muzzin:2013yxa, Ilbert:2013bf, Tomczak:2013bxa}, and $z=0.7$ \protect\citep{Moustakas:2013mp}. Here we see that the JWST-optimized model overpredicts the abundance of lower-mass galaxies relative to low-$z$ measurements. Note that our model was not fit to the data shown in either panel.}
\label{fig:lowz}
\end{figure*}

We evaluate our model---fit to JWST data---to predict HST measurements at a slightly lower redshift ($z=8$). We show the results of this exercise in the left panel of Figure~\ref{fig:lowz}, where our predicted UVLF is compared to the HST data points. We see broad consistency between the model and the HST data for $\Muv\gtrsim -20.5$; the model overpredicts the abundance of brighter galaxies with $\Muv \lesssim -20.5$. This is expected, given that the actual UVLF measurement in the $8.5 < z < 9.5$ redshift range and at $\Muv=-22$ from JWST is about one order of magnitude larger than the HST UVLF reconstruction at $z=8.0$. This suggests a mild tension between the JWST and HST data sets, under the specific model we used in this study.

In the right panel of Figure~\ref{fig:lowz}, we compare our predictions for low-$z$ galaxy abundances to observations of the stellar mass function (SMF), using the collection of observational constraints from \cite{Behroozi_2019}. We find that our model is consistent with the observed SMF at $z=8$, for which we show observations from Spitzer/IRAC data \citep{Song2016}. This agreement is also consistent with the left panel of Figure \ref{fig:lowz}, where we show that our predictions agree with UVLF measurements at $z=8$.

At lower redshifts, we overpredict the abundance of lower-mass galaxies ($M_*\lesssim 10^{10}~M_{\mathrm{\odot}}$) for $z\lesssim 6.0$; this discrepancy grows with decreasing redshift. Since we only fit to high-$z$ JWST galaxies, this is likely a consequence of the JWST preference for rapid star formation. An improvement in the fit could potentially be achieved with a redshift-dependent model of the star formation/outflow rates (e.g., \citealt{Lu240602672}). Meanwhile, at the high-mass end of the SMF, our predictions tend to be more consistent with observations, with a notable exception at $z=2$. We have used the quasar-mode AGN feedback model from \cite{Ostriker10042923}, with a black hole wind luminosity that depends on the product of a black hole wind efficiency and the radiative efficiency of the black hole's accretion disk. The black hole wind efficiency, $\epsilon_\mathrm{w}=2.4\times 10^{-4}$ (comparable to the ``low efficiency'' models considered by \citealt{Ostriker10042923}) at a canonical radiative efficiency of 10\%, is fixed from a calibration to $z\sim 0$ data \citep{Knebe2018}. Thus, our AGN feedback model parameters may need to evolve with redshift to match the $z\sim 2$ data.

\section{Discussion}
\label{sec:discussion}

This work presents a framework for probabilistic modeling of high-$z$ galaxies by leveraging a galaxy-population-synthesis forward model and optimizing it based on directly-observable characteristics of high-$z$ galaxy populations. Here, we outline steps for extending our analysis approach and improving the flexibility of our galaxy evolution model, beyond the relatively simple prescription we have presented in this work.
\begin{itemize}
    \item \emph{Dust contamination.} The astrophysical model we used in this study does not include the effects of dust on the observed galaxy luminosities. UV attenuation by dust can cause intrinsically brighter galaxies to appear fainter; this difference is generally most pronounced in more massive galaxies that have more dust, potentially suppressing the bright end of the UVLF. For example, \cite{Feldmann240702674} predict a shift of $\Delta \Muv \approx 0.5~\mathrm{mag}$ for the brightest high-$z$ galaxies in their hydrodynamic simulations. Including dust would therefore lead our inference to prefer slightly weaker outflow rates, and a stronger outflow rate mass dependence, in order to reproduce the observed UVLF. 
    \item \emph{Bursty star formation.} In the present work, we treat the SFR as an averaged quantity that depends solely on the disk velocity and the amount of gas in a halo. Although a degree of burstiness is captured by our merger trees, hydrodynamic simulations predict a large variability in SFRs that may dwarf the effects of mergers alone (e.g., \citealt{Sun230502713}). Bursty star formation histories yield a large spread in UV luminosity at fixed halo mass, offering an alternative mechanism for increasing the abundance of bright galaxies observed by JWST \citep{Shen230505679}. This mechanism would likely allow for less extreme SFRs and stronger outflow rates than our fiducial inference, and we plan to explore this using an explicit parameterization of short-timescale burstiness in future work. 
    \item \emph{Cosmic variance.} We do not directly account for field-to-field cosmic variance in galaxy clustering, which could impact the interpretation of CEERS and NGDEEP data if, e.g., it includes magnified galaxies in lensed fields with galaxy abundances that are biased relative to the cosmic mean. \cite{Jespersen240300050} show that such cosmic variance introduces super-Poissonian scatter that increases toward higher redshifts. Including cosmic variance in the likelihood would allow models with lower intrinsic abundances of bright galaxies to be consistent with the data, as systems preferentially up-scatter in luminosity with an intrinsic variance that is largest at the bright end. This effect would change the slopes of our predicted UVLFs and therefore would likely weaken constraints on $\alpha_{\mathrm{outflow}}$.
    \item \emph{High-$z$ halo mass function.} As discussed in \cite{Yung230404348}, models commonly adopted for the high-$z$ halo mass function are calibrated to N-body simulations that save relatively few snapshots at these times. Thus, it is possible that the dark matter halo mass functions used in our model (from \citealt{Sheth2001}) are inaccurate at high redshifts.  For the typical halo masses occupied by the high-$z$ galaxies observed by JWST, this inaccuracy is at most $\approx 50\%$, and typically much lower ($\sim$20\%; \citealt{Yung230404348}, Figure 5).
    \item \emph{Initial mass function.} We use a \cite{Chabrier2001} IMF that is calibrated to low-$z$ stellar populations. However, the IMF may be top-heavy at high redshifts. As explored in \cite{Yung230404348}, \cite{Wang2024}, \cite{Cueto2024}, and \cite{Hutter2024}, this would boost the UVLF at early times, which would allow for lower SFRs and higher feedback efficiencies in our model.
    \item \emph{Cosmology.} In addition to galaxy formation physics, the underlying cosmological model can change the predicted population of halos that host galaxies at high redshifts. In particular, enhancement or suppression of the linear matter power spectrum $P(k)$ could affect the UVLF. Using the semi-analytic framework from \cite{Nadler221208584}, we find that the expected $z=0$ descendant mass distribution for the $M_{h}(z=8)\sim 10^{10}~M_{\mathrm{\odot}}$ halos that host high-$z$ JWST galaxies is peaked at a present-day halo mass of $M_{h}(z=0)\sim 5\times 10^{11}~M_{\mathrm{\odot}}$. These systems are sourced by density fluctuations with comoving wavenumbers of $k\sim 2~\mathrm{Mpc}^{-1}$, where masses are associated with wavenumbers in linear theory following \cite{2019ApJ...878L..32N}. Several analyses of high-$z$ JWST galaxies have considered models that affect these scales, including warm dark matter \citep{Maio221103620,Dayal230314239,Lin230605648,Hoeneisen240108730,Liu240413596}, axion dark matter \citep{Bird230710302}, and non-standard primordial power spectra \citep{Parashari230500999,Hirano230611993,Tkachev230713774}. Our framework is well suited to place constraints on these scenarios because it allows us to marginalize over astrophysical uncertainties; we plan to pursue such constraints in future work.
\end{itemize}

\section{Conclusions}
\label{sec:conclusions}

We introduced a framework for population synthesis of high-$z$ galaxies, visualized in Figure~\ref{fig:pipeline}. We further applied probabilistic inference to determine model parameters, using the likelihood function that models data at the level of direct observables: UV apparent magnitudes and redshifts of individual objects. We fit the model to JWST data from the CEERS and NGDEEP surveys, reporting the ranges of star formation and gas outflow rates (and their halo mass dependencies) consistent with JWST galaxy surveys; our results are summarized in Table~\ref{tab:astro_params}, which lists the best-fit parameters and associated uncertainties. 
Our key findings are as follows:
\begin{enumerate}
    \item The upper limit on the star formation timescale is $500~\mathrm{Myr}$ at a disk velocity of $50~\mathrm{km\ s}^{-1}$, at $95\%$ confidence. The characteristic velocity at which the outflow mass loading factor is $\sim 1$ is $150^{+280}_{-60}~\mathrm{km\ s}^{-1}$, also at $95\%$ confidence. We constrain the halo mass dependencies of these effects, and our full posterior distribution reconstruction maps out the degeneracies between these parameters in Figure~\ref{fig:triangle}.
    \item We obtain a good fit to the apparent magnitude--redshift distribution of high-$z$ CEERS and NGDEEP galaxies, and thus their UVLF, in a standard $\Lambda$CDM cosmology (Figures~\ref{fig:muvz_viz}, \ref{fig:uvlf_spread}).
    \item We infer that observable galaxies in CEERS and NGDEEP data at $8.5<z<12$ occupy halos with virial masses of $10^{10\pm 0.5}~M_{\mathrm{\odot}}$ at these redshifts, at $95\%$ confidence (Figure~\ref{fig:mh_pdf}).
    \item Our best-fit model generally predicts faster star formation and weaker feedback than expected from low-$z$ data. As a result, while our predictions are broadly consistent with HST measurements of the UVLF at $z=8$, they overpredict the faint end of the stellar mass function at $z\lesssim 6$. 
\end{enumerate}

The last point could indicate that the star formation and feedback parameters evolve with redshift, or that there may be other astrophysical effects that influence galaxy evolution at lower redshifts, beyond what is captured by our model. Looking forward, we expect that it will be important to model dust, IMF variations, and star formation burstiness in a similar probabilistic framework, and combine low and high-$z$ data in the parameter inference.

We emphasize that our forward-modeling approach is efficient compared to hydrodynamic simulations, and captures more complexity than many empirical models of the high-$z$ galaxy population. Nonetheless, the limited speed of the semi-analytic model evaluations resulted in a pixelized reconstruction of the astrophysical posterior. Thus, it is timely to develop even more efficient forward-modeling methods based on this framework. These tools will ultimately enable a comprehensive understanding of the astrophysics and underlying dark matter physics of galaxies throughout cosmic history.

\section*{Acknowledgements}

We thank Sownak Bose and Peter Behroozi for helpful conversations related to this work. Computing resources used in this work were made available by a generous grant from the Ahmanson Foundation. VG acknowledges the support from NASA through the Astrophysics Theory Program,
Award Number 21-ATP21-0135, the National Science Foundation (NSF) CAREER Grant No. PHY2239205, and from
the Research Corporation for Science Advancement under
the Cottrell Scholar Program. This research was supported in part by grant NSF PHY-2309135 to the Kavli Institute for Theoretical Physics (KITP).


\bibliographystyle{mnras}
\bibliography{ref} 



\appendix

\section{Outflow and Star Formation Rates}\label{app:outflow_sfr}

\begin{figure*}
    \begin{subfigure}[t]{0.49\textwidth}
    \includegraphics[width=\linewidth]{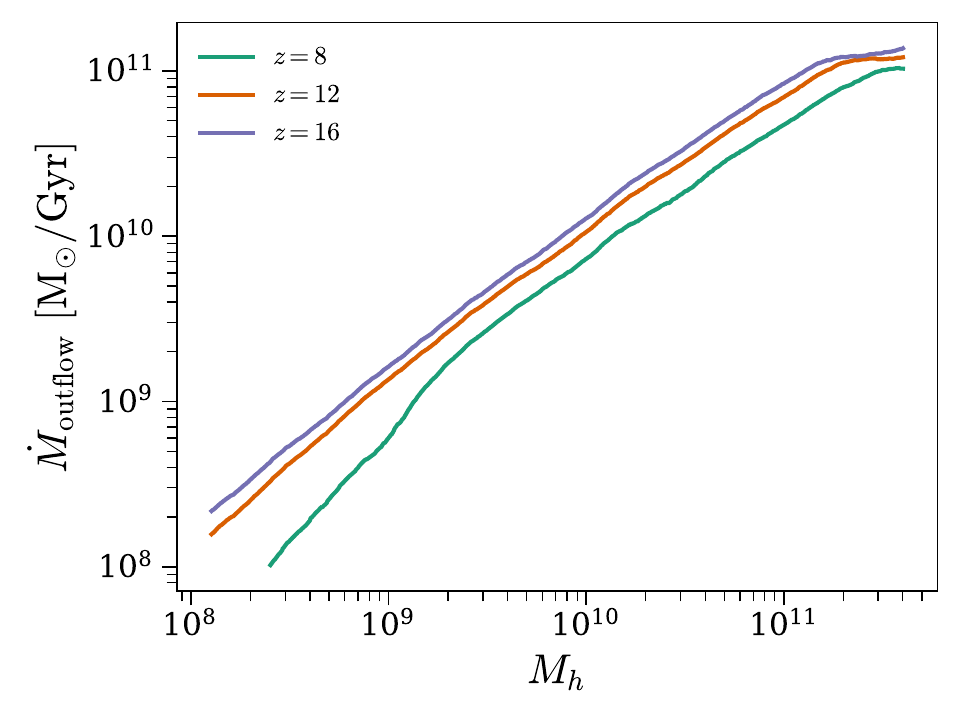}
    \end{subfigure}
    \hfill
    \begin{subfigure}[t]{0.49\textwidth}
    \includegraphics[width=\linewidth]{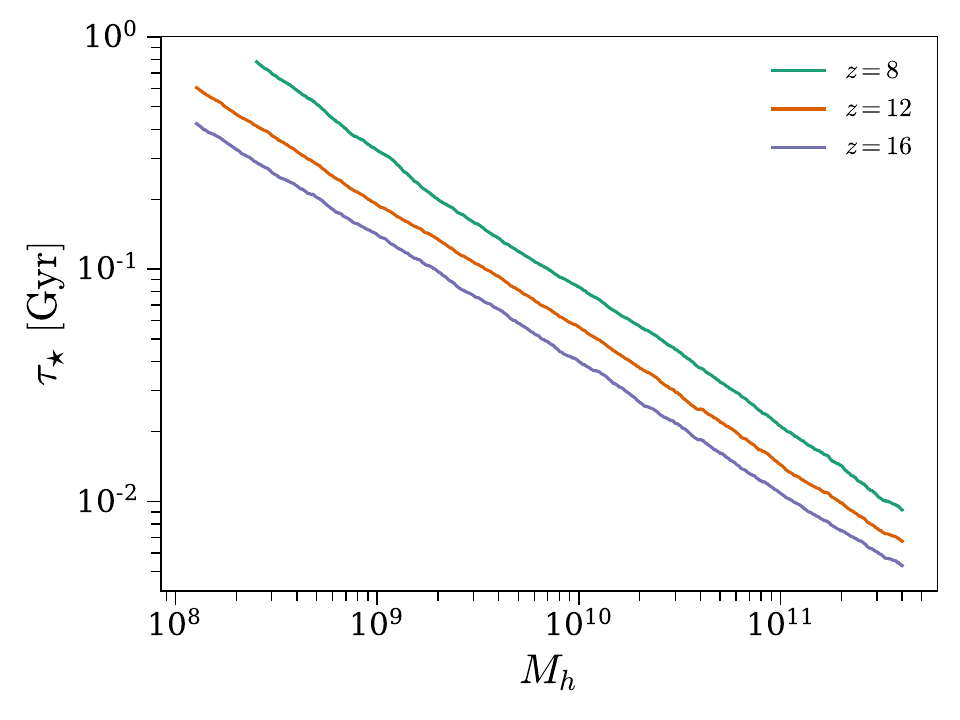}
    \end{subfigure}
    \caption{Average outflow rates $\Mdot_{\mathrm{outflow}}$ and timescale for star formation $\tau_{\star}$ as a function of halo mass for a our best fit model. }
    \label{fig:moutflow_taustar}
\end{figure*}

In Figure~\ref{fig:moutflow_taustar}, we plot the full mass-dependent outflow rate and star formation timescale of our best-fit model at $z=8,12,$ and $16$. Given that the outflow rate and star formation timescale are modeled as power laws, the slope of the lines in the left panel of Figure~\ref{fig:moutflow_taustar} is set by the power-law index parameter $\alphaout$, while the normalization is set by $\Vout$; in the right panel, the slope if set by $\alphastar$ and the normalization is set by $\tau_0$.

\bsp	
\label{lastpage}
\end{document}